\documentclass[twocolumn,showpacs,preprintnumbers,amsmath,amssymb,aps,prl,superscriptaddress]{revtex4-1}
\usepackage{graphicx}
\usepackage{dcolumn}
\usepackage{bm}

\begin{document}

\title{Unified understanding of the valence transition in the rare-earth monochalcogenides under pressure}

\author{I. Jarrige}
\email{To whom correspondence should be addressed: E-mail: jarrige@bnl.gov}
\affiliation{National Synchrotron Light Source II, Brookhaven National Laboratory, Upton, NY 11973-5000, USA}
\author{H. Yamaoka}
\affiliation{RIKEN SPring-8 Center, Sayo, Hyogo 679-5148, Japan}
\author{J.-P. Rueff}
\affiliation{Synchrotron SOLEIL, L'Orme des Merisier, BP 42 Saint-Aubin, 91192 Gif sur Yvette, France}
\affiliation{Laboratoire de Chimie Physique -- Mati\`{e}re et Rayonnement, Universit\'e Pierre et Marie Curie, CNRS, 11 rue Pierre et Marie Curie, 75005 Paris, France}
\author{J.-F. Lin}
\affiliation{Department of Geological Sciences, The University of Texas at Austin, Austin, TX 78712, USA }
\author{M. Taguchi}
\affiliation{RIKEN SPring-8 Center, Sayo, Hyogo 679-5148, Japan}
\author{N. Hiraoka}
\author{H. Ishii}
\author{K.D. Tsuei}
\affiliation{National Synchrotron Radiation Research Center, Hsinchu 30076, Taiwan}
\author{K. Imura}
\affiliation{Department of Physics, Nagoya University, Nagoya 464-8602, Japan}
\author{T. Matsumura}	
\affiliation{Department of Quantum Matter, AdSM, Hiroshima University, Higashi-Hiroshima, 739-8530, Japan }
\author{A. Ochiai}
\affiliation{Center for Low Temperature Science, Tohoku University, Sendai 980-8578, Japan }
\author{H.S. Suzuki}
\affiliation{National Institute for Materials Science, Tsukuba, Ibaraki 305-0047, Japan}
\author{A. Kotani}
\affiliation{RIKEN SPring-8 Center, Sayo, Hyogo 679-5148, Japan}
\affiliation{Photon Factory, Institute of Materials Structure Science, High Energy Accelerator Research Organization, 1-1 Oho, Tsukuba, Ibaraki 305-0801, Japan}

\begin{abstract}
Valence instability is a key ingredient of the unusual properties of $f$ electron materials, yet a clear understanding is lacking as it involves a complex interplay between $f$ electrons and conduction states. Here we propose a unified picture of pressure-induced valence transition in Sm and Yb monochalcogenides, considered as model system for mixed valent $4f$-electron materials. Using high-resolution x-ray absorption spectroscopy, we show that the valence transition is driven by the promotion of a $4f$ electron specifically into the lowest unoccupied (LU) $5d$ $t_{2g}$ band. We demonstrate with a promotional model that the nature of the transition at low pressures is intimately related to the density of states of the LU band, while at high pressures it is governed by the hybridization strength. These results set a new standard for the generic understanding of valence fluctuations in $f$-electron materials.
\end{abstract}


\maketitle
Valence fluctuations play an essential role in some of the $f$-electron systems most exciting behaviors, such as quantum criticality and unconventional superconductivity \cite{yuan,holmes}. But  their understanding  has proven challenging to capture in a  unified theory because they arise from subtle many-body interactions between the $f$ electrons and the conduction states \cite{strange,temmerman}. Pressure on the other hand is an efficient way to act on $f$-electron interactions and localization and thus can serve as a window onto the $f$-electron physics. Notably, the pressure-induced valence transition of the lanthanide monochalcogenides is a paradigmatic illustration of $f$ delocalization phenomena and their tremendous diversity. For instance, SmS undergoes a first-order transition to near-trivalency coinciding with the onset of magnetic ordering \cite{annese1}, whereas the transition of YbS into an intermediate-valent correlated metal is extremely sluggish \cite{matsunami}.
These diverse behaviors, while establishing a severe test bed for theoretical understanding of $f$-electron systems, have been overlooked for the past decades. Here, we address them in the light of a direct measurement of the electronic structure of Sm and Yb monochalcogenides (SmS, SmSe, SmTe, YbS, YbSe) under pressure performed using high-resolution x-ray absorption spectroscopy in the partial fluorescence yield (PFY-XAS) mode. Our data reveal that the valence discontinuity and concomitant closing of the gap under pressure are caused by the specific filling of the lowest unoccupied (LU) $5d$ $t_{2g}$ band by a $4f$ electron, offering a unified picture of electron delocalization and intermediate valency. Using a promotional model, we show that both the steepness and the amplitude of the valence transition of the Sm compounds increase with the density of states (DOS) of this LU band. When exceeding a critical DOS, the transition becomes first order for SmS. The model fails to describe the prolonged transition of the Yb compounds, which suggests that at high pressures the valence transition is impeded by enhanced hybridization. Ultimately, we suggest that our analysis can serve as a prototype for understanding the general mechanism of delocalization in $f$ electron materials. 

The PFY-XAS measurements were carried out at room temperature at the Taiwan beamline BL12XU, SPring-8. Details of the experimental setup have been published elsewhere \cite{yamaoka1}. For each compound, a fragment of single crystal of $\sim$100 $\mu$m in size was loaded into the sample chamber of a Be gasket with silicone oil used as pressure transmitting medium. Pressure was achieved using a diamond anvil cell.
The pressure dependence of the Sm $L_{3}$ edge measured on the Sm monochalcogenides by PFY-XAS is shown in the top panels of Fig.~1 (a). The spectra consist of two main peaks corresponding to the dipolar transition $2p\rightarrow5d$ for the divalent and trivalent states, respectively decreasing and increasing under pressure. A closer look at the leading edge of the divalent peak reveals a shoulder located around 2.3 (SmS), 2.0 (SmSe), and 1.6 (SmTe)  eV below the peak maximum (open circle in Fig.~1(a)). A fit of the divalent component actually requires the use of three Gaussian functions in order to reproduce the low-energy shoulder, the main peak, and a high-energy shoulder, hereafter referred to as A, B, and C (cf. Fig.~1). 

\begin{figure}
\includegraphics[bb=280 147 700 575,clip=,width=11.5cm,angle=0] {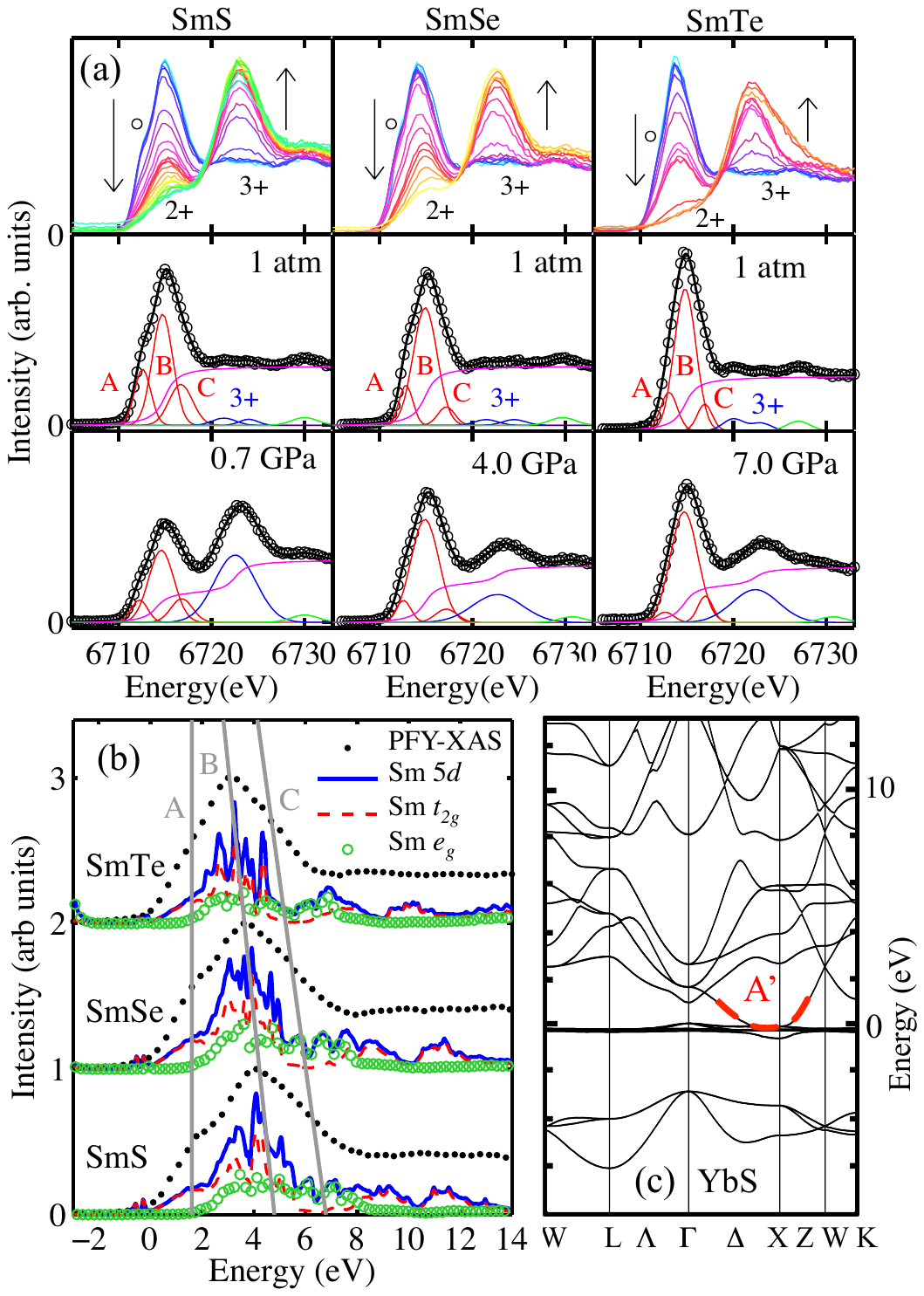}
\caption{(Color online) (a) Pressure dependence of the Sm-$L_{3}$ PFY-XAS spectra for SmS, SmSe, SmTe (top row), and examples of fit performed on spectra both before (center row) and after (bottom row) collapse. Three components A, B, C, are used to fit the 2+ feature. Arrows indicate the direction of increasing pressure. (b) Sm $5d$, $t_{2g}$, $e_{g}$ DOS calculations for SmS, SmSe, SmTe compared with the corresponding PFY-XAS spectra. As indicated by the solid lines, the energy shift between the peaks A, B, and C decreases across the series SmS, SmSe, SmTe. (c) Band structure of YbS. The low-energy, isolated part of the band A, A', is highlighted.}
\end{figure}

\begin{figure}
\includegraphics[width=.45\textwidth] {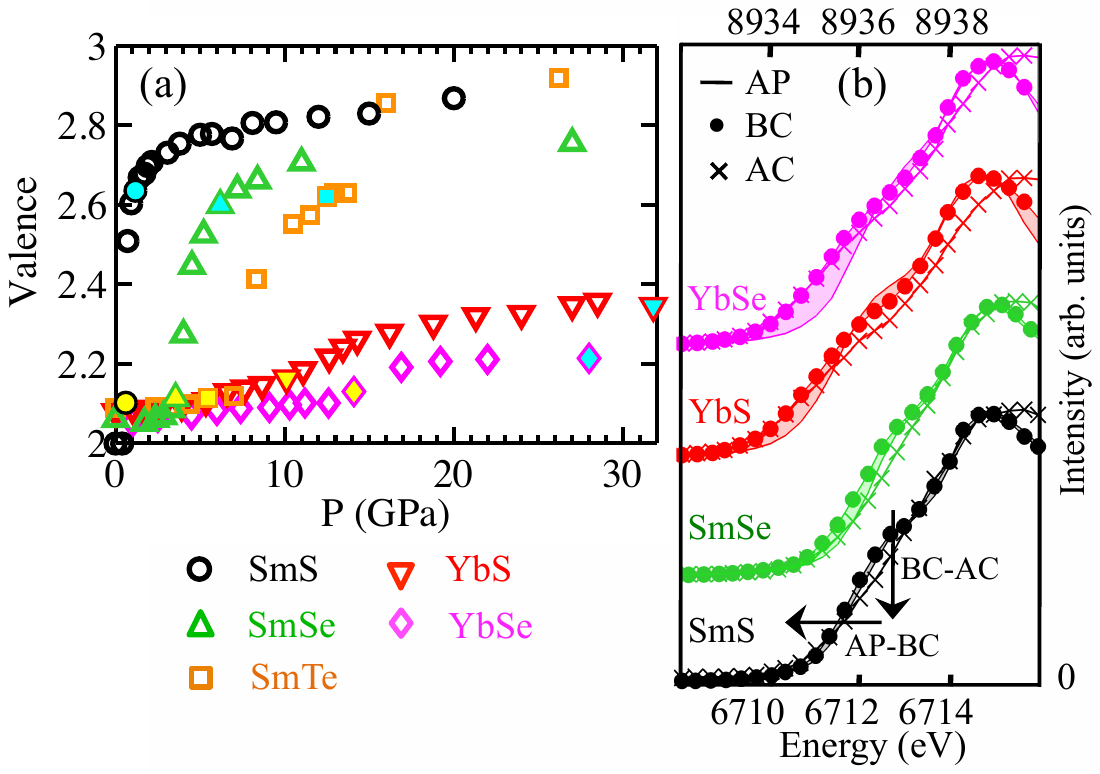}
\caption{(Color online) (a) Pressure dependence of the valence. Yellow and cyan filled markers respectively correspond to before collapse (BC) and after collapse (AC). (b) Enlarged divalent region of three spectra per compound normalized in intensity to the maximum of B, measured at ambient pressure (AP), BC, and AC. The arrows illustrate the low-energy shift of the shoulder A between AP and BC, and its collapse between BC and AC.}
\end{figure}

In order to understand the origin of these spectral features, we have calculated the electronic structure of the Sm and Yb monochalcogenides using the density functional theory within the generalized gradient approximation (GGA). The DOS obtained for the Sm $5d$ band and the crystal-field split $t_{2g}$ and $e_{g}$ subbands are compared with the experimental spectra in Fig.~1(b) for the Sm compounds. We note that the Coulomb interaction is not taken into account and therefore the calculated bands are metallic. The structures A and C are found to be respectively of pure $t_{2g}$ and $e_{g}$ character, while B is a mix of both. The energy splitting between the three structures and the relative intensity of A and C increase along the series SmTe $<$ SmSe $<$ SmS, which reflects the increase of the crystal field as the lattice constant decreases. We can therefore conclusively rule out the previous assignment of A to the quadrupolar pre-edge \cite{annese1,annese2}. Also, the band structures calculated for YbS in Fig.~1(c) and for the Sm compounds in ref. \cite{antonov1} show that the low-energy part of A, which we call A', corresponds to an isolated $t_{2g}$ band sandwiched between the Fermi level and the rest of the conduction band, making it the LU band. 

The pressure dependence of the valence $v$ as derived from the fits of the PFY-XAS spectra is shown for the five monochalcogenides in Fig.~2(a). In previous estimations of $v$, A was incorrectly assigned to a quadrupolar-like feature and therefore not contributing to the divalent weight, which resulted in an overestimation of $v$ by $\sim$~0.1 for SmS and 0.3 for YbS \cite{annese1,annese2}. All five compounds display a steady increase in $v$ at their insulator-to-metal transition pressure, coinciding with the pressure ranges reported for their volume collapse \cite{lebihan}. The abrupt increase in $v$ around 15 GPa for SmTe is due to a structural transition. The onset pressure of the collapse gradually increases as a function of both the ligand (S$<$Se$<$Te) and more starkly the rare-earth (Sm$\ll$Yb). This reflects the difference in the gap values at ambient pressure, $\varepsilon_{gap}$ = 0.20 eV for SmS, 0.45 eV for SmSe, 0.65 eV for SmTe, 1.4 eV for YbS, and 1.75 eV for YbSe~\cite{jayaraman,wachter,syassen,suryanarayanan}, and the idea that the wider the gap, the larger the pressure to close it.

We now address the large variety of slopes and amplitudes of the transitions, and its intimate connection with the LU band. We start by showing in Fig.~2(b) the enlarged divalent region of three PFY-XAS spectra selected for four compounds, SmS, SmSe, YbS, YbSe, normalized in intensity to the maximum of B. These spectra were collected at ambient pressure (AP), just before the collapse (BC), and after the collapse (AC). The BC and AC pressures are respectively indicated by yellow and cyan filled markers in Fig.~2(a). The changes brought on by pressure on the lineshape of A are significant. First, judging from the shape of the shaded area between the AP and BC spectra, A stretches towards low energies between AP and BC, and the extent of the stretch increases with the BC pressure $P_{\rm BC}$, i.e. SmS$<$SmSe$<$YbS$<$YbSe. Second, the collapse coincides with a decrease of A, implying that A gets filled during the electronic transition \cite{si}.

In Fig.~3(b), the pressure dependence of key parameters derived from the fits of the PFY-XAS spectra gives a more precise view of the transition mechanism. For A, we consider the lower-energy portion of the peak which does not overlap with B, A', and the higher-energy portion which does, A'' (cf. Fig. 3(a)), in order to distinguish the filling of the low-energy part of the LU band truly isolated from the rest of the conduction band (A') from that of the higher-energy, hybridized part of the LU band (A''). The sum I$_{\rm A}$+I$_{\rm B}$+I$_{\rm C}$ is scaled to the total number of $5d$ holes, i.e. between 10 for $v=2$ and 9 for $v=3$. The pressure axis is rescaled by $P_{\rm BC}$ indicated by the vertical dashed lines. 
The error on the intensities and width is estimated to be under $\pm 5\%$ \cite{si}. Confirming the trend observed in Fig.~2(b), a close correspondence is observed between the increase of $v$ and the decrease of A. More strikingly for Sm, the pressure ranges of the successive decline of I$_{\rm A'}$ and I$_{\rm A''}$ correlate respectively with the valence jump and the start of the slow increase (cf.\ blue circles and black crosses in Fig.~3(b)). This demonstrates that the volume collapse transition results from the dumping of $f$ electrons into A', while the transition notably slows down as soon as $f$ electrons are promoted into A'', and slows even more when they start to fill the B band beyond AC. For YbS and YbSe the whole transition is considerably slower, as A' is not yet filled up at 30 GPa. The increase in I$_{\rm B+C}$ observed across the transition for all five compounds occurs as a result of B significantly broadening, causing portions of A and C to merge with B \cite{si}. The amplitude of the valence discontinuity is therefore determined by the remaining intensity of A. Also noteworthy is the finding that the monochalcogenide gaps are closed by a combination of both band broadening and shift under pressure, as seen from the $E_{\rm A}$ and $W_{\rm A}$ panels.

\begin{figure*}
\begin{center}
\includegraphics[width=.95\textwidth] {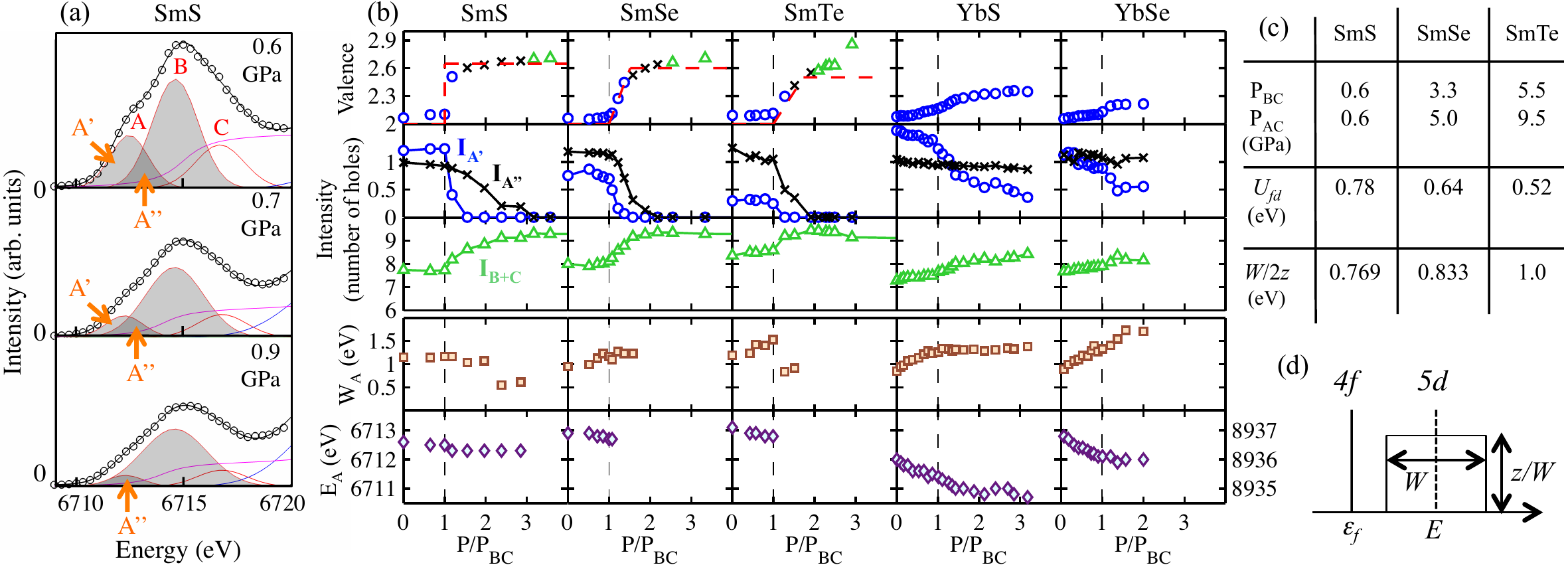}
\end{center}
\caption{(Color online) (a) Illustration of the decomposition of A into A' (light grey) and A'' (dark grey). (b) Pressure dependence of $v$ estimated experimentally (markers) and with the model (dashed line) (top row), and of the intensity of the peaks A', A'', B+C (respectively second and third row from top). The use of different markers highlights the correspondence between the $v$ transition and the filling of A', A'', and B+C for the Sm compounds. Pressure dependence of the width and energy of the peak A (respectively second and first row from bottom). The vertical dashed lines indicate BC pressure, P$_{BC}$. (c) Parameters of the promotional model. (d) Schematic illustration of the promotional model.}
\end{figure*}

For a systematic understanding, we have developed a simple model of the promotion of an electron from a $4f$ level of width zero into a square $5d$ band (corresponding to the LU band) of width $W$ and occupancy $z$, as illustrated in Fig.~3(d). $W$ is taken as 1 eV, which is an approximation of $W_{A}$ at BC (cf. Fig. 3(b)), and $z$ is taken as $v_{AC}$-2. We consider that $U_{fd}$, the Coulomb interaction between the $4f$ and the conduction electrons, is effective. For $U_{fd} < \frac{W}{2z}$ the valence transition is continuous, and its pressure range, defined by the values of $P_{\rm BC}$ and $P_{\rm AC}$, depends on the value of $U_{fd}$. In this case, the energy of the Fermi level $\varepsilon _{F}$ is given by \cite{si}:
\begin{equation}
\varepsilon _{F}=\frac{\varepsilon _{f}-2U_{fd}(\frac{z}{W})(E-\frac{W}{2})}{1-2U_{fd} (\frac{z}{W})},
\end{equation}
\noindent where $\varepsilon _{f}$ is the energy of the $4f$ level and $E$ is the center of the square $5d$ band. The valence is then expressed as:
\begin{equation}
v=2-\frac{z}{W} \bigg[\frac{E-\varepsilon _{f}-\frac{W}{2}}{1-2U_{fd}(\frac{z}{W})}\bigg].
\end{equation}
In our calculation of $v$, we put
\begin{equation}
E-\varepsilon_f-\frac{W}{2} = \varepsilon_{gap}-cP,
\end{equation}
with $c = \varepsilon_{gap}/P_{\rm BC}$.

\indent For $U_{fd} > \frac{W}{2z}$, at a critical value of the pressure there is a first-order transition and the valence changes discontinuously from 2 to 2+$z$. The critical pressure is determined based on the condition that the total energy of the system under pressure is equivalent to the value at ambient pressure.

The transition for the Sm compounds is best reproduced for the parameters indicated in Fig.~3(c). Remarkably enough, the main features of the valence transition are properly captured for the three Sm compounds, a higher DOS of the LU band ($\frac{z}{W}$) resulting in a steeper transition, ultimately first order for SmS, as shown in the top panel of Fig.~3(b). This lends support to the relevance of the promotional model, which has been an open question since the work of Ramirez and Falicov on Ce \cite{ramirez}. Because our model describes the linear filling of one band, it approximates a constant slope through the transition between BC and AC, and does not account for the filling of A'' and B beyond AC. For YbS and YbSe, our model yields unphysical negative values of $U_{fd}$. For YbS for instance, using in Eqs. (2) and (3) the experimental values $P_{\rm BC}$ = 8.7 GPa, $P_{\rm AC}$ = 18.0 GPa, $z$ = 0.4, and $\varepsilon_{gap}$ = 1.4 eV, together with $W$ = 1.0 eV, we obtain $U_{fd}$=-0.62 eV. A very plausible explanation is that hybridization effects between the $4f$ electron and the $5d$ band become stronger at the higher transition pressures of the Yb compounds, so that A' no longer retains its isolated character.

This analysis thus delineates two broad scenarios for the electronic transition of the monochalcogenides. In the first, the LU band is isolated from the rest of the conduction band, allowing for a steep transition, whose slope and amplitude are proportional to the DOS of the LU band. Hybridization thrives in the second scenario, which leads to a sluggish transition, for which a Kondo description may be appropriate. We note that because $\frac{z}{W}$ decreases and hybridization increases with increasing pressure, slower transitions may be more often found at higher pressures. This approximation holds for other rare-earth monochalcogenides, the transition being first order for TmSe$_{0.45}$Te$_{0.55}$ at 1 GPa \cite{neuenschwander}, second order but steep for TmTe at 2 GPa\cite{jarrige}, and sluggish between 16 and 20 GPa for EuSe~\cite{syassen2}. Our interpretation may also be generalized to other divalent $4f$ materials. EuO, for instance has an A shoulder in its $L_{3}$ spectrum, and a sluggish transition was recently reported above 14 GPa \cite{souza}. On the other hand divalent $4f$ materials without an A shoulder show a sluggish transition which starts as soon as above 1 atm, see for instance Pr \cite{bradley}, Eu \cite{rohler}, Yb \cite{fuse}, YbAl$_{2}$ \cite{dallera}, YbGa$_{1.15}$Si$_{0.85}$ \cite{yamaoka1}, or Cd$_{5.7}$Yb \cite{kawana}. We believe this is because $f$ electrons are directly promoted into the main $5d$ conduction band.

We have presented a simple framework which allows for a unified description of the diverse pressure-induced electronic phase transitions of Sm and Yb monochalcogenides, with implications for the $4f$ delocalization phenomenon in general. We have shown that the transition corresponds to the filling of the LU band. This band is at the heart of the diverse electronic transitions of the monochalcogenides, as the slope and amplitude of the transition depend on both its DOS and degree of hybridization. Our analysis can also seemingly describe the delocalization in a variety of other compressed $4f$ electron materials. It may also be extended to the $5f$ electron systems, which have a broad $6d$ unoccupied band and usually show protracted electronic transitions under pressure \cite{rueff}.

\begin{acknowledgments}
One of the authors (A. Kotani) was partially supported by a Grant-in-Aid for Scientific Research C (No. 90029504) from the Japan Society for the Promotion of Science.
\end{acknowledgments}


\begin{references}
\bibitem{yuan}H. Q. Yuan, F. M. Grosche, M. Deppe, C. Geibel, G. Sparn, and F. Steglich, Science {\bf 302}, 2104 (2003).
\bibitem{holmes} A. T. Holmes, D.  Jaccard, and K. Miyake, J. Phys. Soc. Jpn. {\bf 76}, 051002 (2007).
\bibitem{strange} P. Strange, A. Svane, W. M. Temmerman, Z. Szotek, and H. Winter, Nature {\bf 399}, 756 (1999).
\bibitem{temmerman} W. M. Temmerman, A. Svane, L. Petit, M. L\"{u}ders, P. Strange, and Z. Szotek, Phase Transitions {\bf 80}, 415 (2007).
\bibitem{annese1} E. Annese, A. Barla, C. Dallera, G. Lapertot, J.-P. Sanchez, and G. Vank\'{o}, Phys. Rev. B {\bf 73}, 140409(R) (2006).
\bibitem{matsunami} M. Matsunami, H. Okamura, A. Ochiai, and T. Nanba, Phys. Rev. Lett. {\bf 103}, 237202 (2009).
\bibitem{yamaoka1} H. Yamaoka {\it et al.}, Phys. Rev. B {\bf 83}, 104525 (2011).
\bibitem{annese2} E. Annese, J.-P. Rueff, G. Vank\'{o}, M. Grioni, L. Braicovich, L. Degiorgi, R. Gusmeroli, and C. Dallera, Phys. Rev. B {\bf 70}, 075117 (2004).
\bibitem{antonov1} V.N. Antonov, B.N. Harmon, and A.N. Yaresko, Phys. Rev. B {\bf 66}, 165208 (2002).
\bibitem{lebihan} T. Le Bihan, S. Darracq, S. Heathman, U. Benedict, K. Mattenberger, and O. Vogt, J. Alloys Compd. {\bf 226}, 143 (1995).
\bibitem{jayaraman} A. Jayaraman, V. Narayanamurti, E. Bucher, and R. G. Maines, Phys. Rev. Lett. {\bf 25}, 1430 (1970).
\bibitem{wachter} P. Wachter, Handbook on the Physics and Chemistry of Rare Earths {\bf 19}, 177 (1994).
\bibitem{syassen} K. Syassen, H. Winzen, H.G. Zimmer, H. Tups, and J.M. Leger, Phys Rev B {\bf 32}, 8246 (1985).
\bibitem{suryanarayanan} R. Suryanarayanan, J. Ferr\'e, and B. Briat, Phys. Rev. B {\bf 9}, 554 (1974).
\bibitem{si} See supplementary material for details.
\bibitem{ramirez} R. Ramirez and L.M. Falicov, Phys. Rev. B {\bf 3}, 2425 (1971).
\bibitem{neuenschwander} J. Neuenschwander and P. Wachter, Phys. Rev. B {\bf 41}, 12693 (1990).
\bibitem{jarrige} I. Jarrige, J.-P. Rueff, S. R. Shieh, M. Taguchi, Y. Ohishi, T. Matsumura, C.-P. Wang, H. Ishii, N. Hiraoka, and Y.Q. Cai, Phys. Rev. Lett. {\bf 101}, 127401 (2008).
\bibitem{syassen2} K. Syassen, Physica {\bf 139/140B}, 277 (1986).
\bibitem{souza} N. M. Souza-Neto, J. Zhao, E. E. Alp, G. Shen, S. V. Sinogeikin, G. Lapertot, D. Haskel,  arXiv:1024.3003. 
\bibitem{bradley} J. A. Bradley, K. T. Moore, M. J. Lipp, B. A. Mattern, J. I. Pacold, G. T. Seidler, P. Chow, E. Rod, Y. Xiao, and W. J. Evans, Phys. Rev. B {\bf 85}, 100102(R) (2012).
\bibitem{rohler} J. R\"ohler, Physica B {\bf 144}, 27 (1986).
\bibitem{fuse} A. Fuse, G. Nakamoto, M. Kurisu, N. Ishimatsu, and H. Tanida, J. Alloys Compd. {\bf 376}, 34 (2004).
\bibitem{dallera} C. Dallera, E. Annese, J.-P. Rueff, A. Palenzona, G. Vank\'{o}, L. Braicovich, A. Shukla, and M. Grioni, Phys. Rev. B {\bf 68}, 245114 (2003).
\bibitem{kawana} D. Kawana, T. Watanuki, A. Machida, T. Shobu, K. Aoki, and A.P. Tsai, Phys. Rev. B {\bf 81}, 220202(R) (2010).
\bibitem{rueff} J.-P. Rueff, S. Raymond, A. Yaresko, D. Braithwaite, Ph. Leininger, G. Vank\'{o}, A. Huxley, J. Rebizant, and N. Sato, Phys. Rev. B {\bf 76}, 085113 (2007).
\end{references}
\end{document}


\title{Supplementary Material:
Unified understanding of the valence transition in the rare-earth monochalcogenides under pressure}

\author{I. Jarrige}
\author{H. Yamaoka}
\author{J.-P. Rueff}
\author{J.-F. Lin}
\author{M. Taguchi}
\author{N. Hiraoka}
\author{H. Ishii}
\author{K.D. Tsuei}
\author{K. Imura}
\author{T. Matsumura}	
\author{A. Ochiai}
\author{H.S. Suzuki}
\author{A. Kotani}

\maketitle

\section{Data analysis}
The pressure dependence of the Yb $L_{3}$ edge measured on YbS and YbSe by PFY-XAS is shown in the top panels of Fig.~S1. Examples of fits for both compounds are shown for the lowest and highest measured pressures in the center and bottom panels, respectively. 

The decrease of the shoulder A through the pressure-induced volume collapse, shown in Fig.~2(b), is further illustrated in Fig.~S2. In this figure we have superimposed representative spectra measured at four successive pressures across the collapse with the spectrum measured before the collapse for SmS and SmSe. All spectra are normalized in intensity to the maximum of the B feature. The shoulder A is seen to gradually decrease across the collapse. In Fig.~S3 we show the effect on the fit of the $\pm5\%$ error which we consider for I$_{A}$ and W$_{A}$.

Fig.~S4 shows the increase of $W_{B}$ across the collapse. This broadening causes an increase of the overlap of B with A and C. We attribute the concomitant decrease of A and C illustrated in Fig.~3(a) for SmS to this enhanced mixing, which results in the merging of part of A and C with B.

\begin{figure}[H]
\centering
\includegraphics[width=.3\textwidth] {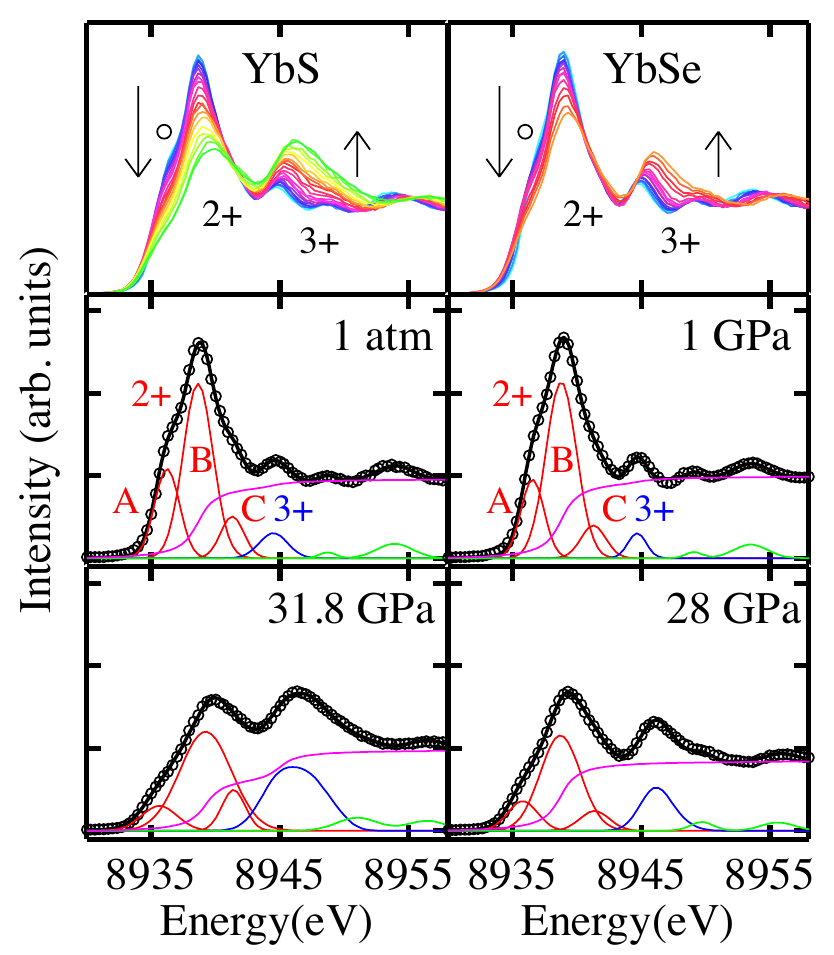}
\caption{(Color online) Pressure dependence of the Yb-$L_{3}$ PFY-XAS spectra measured on YbS and YbSe (top row), and examples of fit performed on the spectra measured at the lowest (central row) and highest pressure (bottom row) for YbS (left column) and YbSe (right column). Three peaks, A, B, C, are used to fit the 2+ feature. In the top-row panels the arrows indicate the direction of increasing pressure.}
\end{figure}

\begin{figure}[H]
\centering
\includegraphics[width=.28\textwidth] {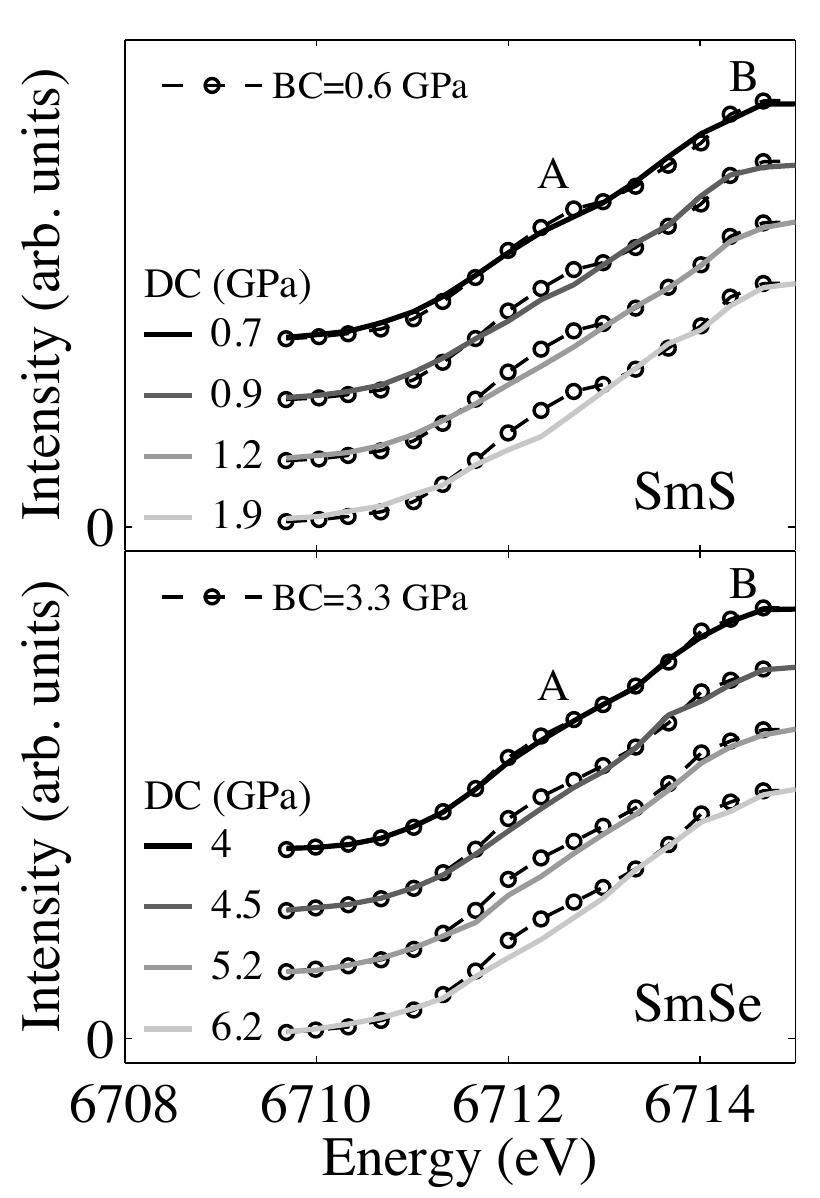}
\caption{Evolution of the feature A across the pressure-induced valence transition in SmS (top panel) and SmSe (bottom panel). For each compound, the Sm-$L_{3}$ PFY-XAS spectrum measured just before collapse (BC, dashed line and circles) is overlapped with four spectra measured during collapse (DC, solid line). All spectra are normalized in intensity to the maximum of the B feature.}
\end{figure}

\begin{figure}
\centering
\includegraphics[width=.4\textwidth] {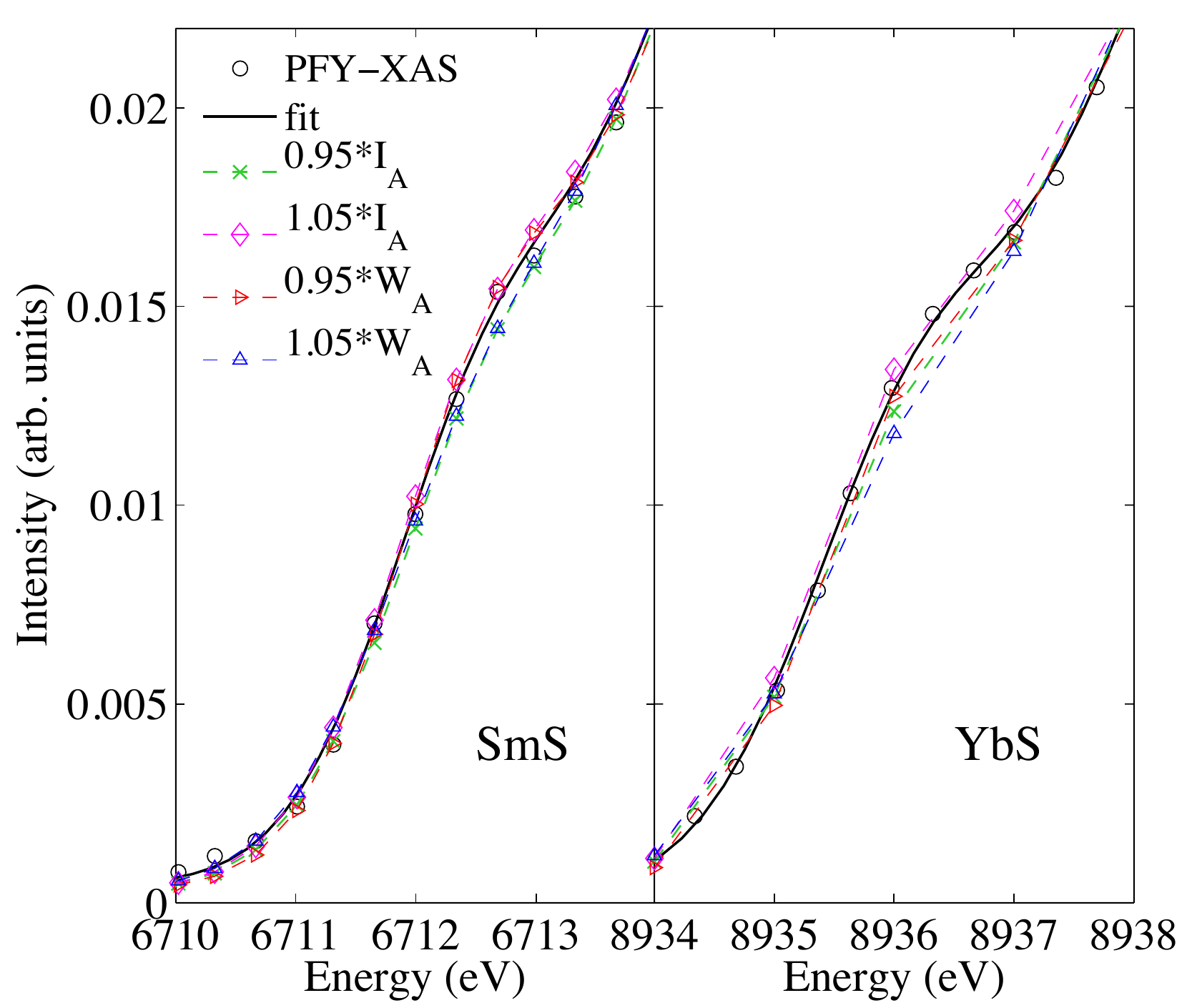}
\caption{(Color online) Illustration of the uncertainty on the fit around the region of the shoulder A for SmS (left panel) and YbS (right panel). Compared with the PFY-XAS spectrum (circles) are the fit (solid line), the fit where the as-derived value of $I_{A}$ was decreased (dashed line and crosses) or increased (dashed line and diamonds) by 5\%,  and also the fit where the as-derived value of $W_{A}$ was decreased (dashed line and right-pointing triangles) or increased (dashed line and up-pointing triangles) by 5\%.}
\end{figure}

\begin{figure}[b]
\begin{center}
\includegraphics[width=.5\textwidth] {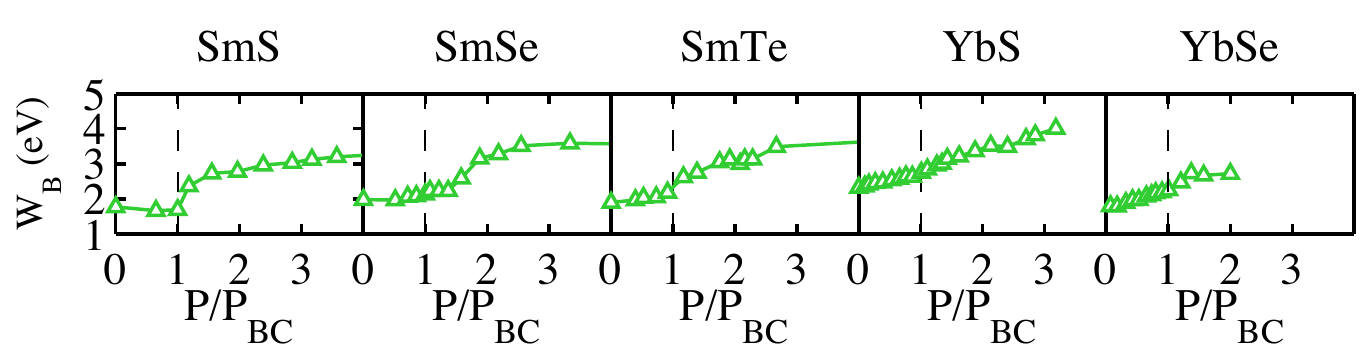}
\end{center}
\caption{Pressure dependence of the full width at half maximum of the peak B, $W_{B}$, derived from the fits for all five compounds.}
\end{figure}

\section{Model}
We first explain how the equation (1) in the article is derived. If we assume that $n$ $4f$ electrons per Sm are promoted to the $5d$ band in the system sketched in Fig. 3(d), the value of $n$, and therefore of the Fermi level $\varepsilon_{F}$, is determined for the stable state by minimizing the total energy change $E_{band}+E_{4f}+E_{int}$ with respect to $n$ (or $\varepsilon_{F}$):\\
\begin{equation}
\frac{d}{d\varepsilon_{F}}(E_{band}+E_{4f}+E_{int})=0,
\end{equation}
where $E_{band}$ and $E_{4f}$ are respectively the kinetic energy changes of the electrons in the $5d$ band and of the $4f$ electrons per Sm, and $E_{int}$ the Coulomb interaction energy between the $5d$ band electrons and the $4f$ holes per Sm:\\
\begin{equation}
E_{band}=\frac{z}{2W}\bigg[\varepsilon_{F}^{2}-\bigg(E-\frac{W}{2}\bigg)^{2}\bigg],
\end{equation}
\begin{equation}
E_{4f}=\bigg[1-\frac{z}{W}\bigg(\varepsilon_{F}-E+\frac{W}{2}\bigg)\bigg]\varepsilon_{f},
\end{equation}
\begin{equation}
E_{int}=-U_{fd}\bigg[\frac{z}{W}\bigg(\varepsilon_{F}-E+\frac{W}{2}\bigg)\bigg]^{2},
\end{equation}
with $W$ and $z$ being respectively the width and the occupancy in the post-collapse state of the $5d$ band, and $U_{fd}$ the Coulomb interaction between the $4f$ and the conduction electrons.

Next, the expression for the valence in the equation (2) of the article is obtained by:\\

$v=2+n$,\\
with $\varepsilon_{F}$ and $n$ being connected by the following relation:\\
\begin{equation}
n=\frac{z}{W}\bigg[\varepsilon_{F}-E+\frac{W}{2}\bigg].
\end{equation}